# Ab-initio NEGF Perspective of Ultra-Scaled CMOS: From 2D-material Fundamentals to Novel Dynamically-Doped Transistors.

*Aryan Afzalian*.*

imec, Kapeldreef 75, 3001 Leuven, Belgium

Email: Aryan.Afzalian@imec.be

ABSTRACT. Using accurate dissipative DFT-NEGF atomistic-simulation techniques within the Wannier-Function formalism, we give a fresh look at the possibility of sub-10 nm scaling for high-performance CMOS applications. We show that a combination of good electrostatic control together with a high mobility is paramount to meet the stringent roadmap targets. Such requirements typically play against each other at sub-10 nm gate length for MOS transistors made of conventional semiconductor materials like Si, Ge or III-V and dimensional scaling is expected to end around 12 nm gate-length. We demonstrate that using alternative 2D channel materials, such as the less explored $HfS_2$ or $ZrS_2$, high-drive current down to about 6 nm is, however, achievable. We also propose a novel transistor concept, the Dynamically-Doped

Field-Effect Transistor, that scales better than its MOSFET counterpart. Used in combination with a high-mobility material such as $HfS_2$, it allows for keeping the stringent high-performance CMOS on current and competitive energy-delay performance, when scaling down to 1 nm gate length using a single-gate architecture and an ultra-compact design. The Dynamically-Doped Field-Effect Transistor further addresses the grand-challenge of doping in ultra-scaled devices and 2D materials in particular.

Scaling and Moore's law, that sets the footprint area of a transistor to scale by a factor 2, that is the transistor gate length $L$ to scale by a factor $\sqrt{2}$, every 2 years, have been the driving force of the electronic industry.[1] Today $L$ has been scaled well below 20 nm and further scaling has become increasingly difficult due to short-channel effects (SCE) that degrade the subthreshold slope (SS) of a transistor (i.e., the efficiency with which the current is switched from off to on state by changing the gate bias). SCE lead to an increased off-state leakage current, $I_{OFF}$. To mitigate SCE, i.e., to keep a good electrostatic control of the gate over the transistor channel, its thickness, $t_s$, has to be scaled as well.[1,2] Also, transistors have evolved from planar single-gate transistors to 3D multi-gate devices such as FINFETs[3], nanowires[4,5] and nanosheets.[6] As a rule of thumb, in a multi-gate device, the channel thickness $t_s$ has to be of the order of ½ $L$ in order to keep the electrostatic integrity leading to $t_s$ of a few nm only in modern advanced nanoscale technologies.[7] At such value of $t_s$, conventional "3D" semiconductors, like Si, or possible high-mobility channel-replacement materials like Ge[3] or III-V,[5] suffer from quantum-confinement (QE) effects that strongly deteriorate their performance (e.g., current drive, gate coupling, mobilities...),[8,9,10] as well as, lead to increased variability (e.g., strong threshold-voltage variations with surface roughness for instance). [2,10,11] It is commonly accepted that conventional dimensional scaling will stop for $L$ of the order of 10 nm. The current international roadmap for device and system (IRDS) predictions have actually forecast that gate-length scaling will stop for a $L$ of 12 nm.[12]

As an attempt to further push the scaling, transistors made of novel 2D materials,[13] i.e., an atomistically thin layer of material that does not create strong atomic bonds in the 3rd dimension, such as transition-metal dichalcogenide (TMD)[14,15,16] or black phosphorus ($P_4$),[14,17] are being actively investigated as future replacement of Si as channel material. These materials would offer the ultimate electrostatic control and are free from the quantum confinement due to their 2D nature. In principle also, their thicknesses could be well controlled which would remove the variability issue. The research development on 2D material is still at an early stage today. Despite the ever-growing list, including several thousands of newly discovered such materials,[18] an ideal CMOS candidate for sub-10nm channel has not yet emerged. Experimentally, only a subset of 2D-material transistors, such as those using $MoS_2$, $WS_2$, $WSe_2$ and $P_4$,[14,15,17] have been explored and the current drive is typically too low for high-performance (HP)-CMOS applications.[12] Although the low drive-current is, at least in part, related to the immaturity of the technology, the fundamental physics and performance of these transistors is not yet fully elucidated. Even using 2D materials, scaling $L$ below 10 nm is further complicated by an additional quantum-mechanical short-channel effect. This effect, called source-to-drain tunneling (SDT), which is the ability of the electronic quantum-mechanical waves to evanescently leak through the channel barrier, further degrades $SS$ and $I_{OFF}$.

Here, using our state-of-the-art DFT-NEGF ATOmistic MOdelling Solver (ATOMOS),[19] we offer unique insights on several 2D-material physics and performance, including those of the less explored $HfS_2$ and $ZrS_2$ that feature appealing performance for ultra-scaled CMOS. We demonstrate the possibility of $L = 6$ nm high-performance devices, providing that high doping can be achieved. Finally, we demonstrate that further geometrical scaling, down to a 1 nm gate length footprint, is possible using a new device concept, the Dynamically-Doped Field-Effect Transistor ($D_2$-FET). The $D_2$-FET concept further addresses the difficult challenge of doping in nanoscale devices[12] and 2D materials in particular,[20] and the need for chemical doping could be suppressed.

## RESULTS AND DISCUSSIONS

### **Device structure and methodology:**

The schematic of the studied monolayer (1ML) double-gate (DG) MOSFETs is shown in Fig. 1. An intrinsic channel of length $L$ is used. The source- and drain- (S&D) extension regions are doped with a concentration $N_{SD}$. An abrupt junction profile is assumed. The 2nm thick $HfO_2$ gate oxide has a relative permittivity $\varepsilon_R$ = 15.6 and an equivalent oxide thickness EOT = 0.5 nm. The work function of the gate-voltage bias, $V_G$, is typically adjusted to achieve a fixed $I_{OFF}$ value at $V_G$ = 0V. A low-K spacer oxide with $\varepsilon_R$ = 4 surrounds the S&D extensions. Ohmic contacts are assumed with S&D bias $V_S$ = 0 V and $V_D$ respectively.

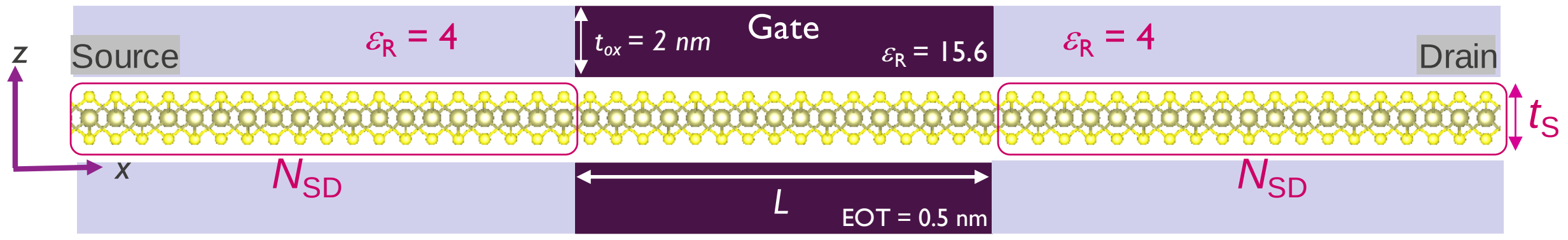

Figure 1. Schematic view of a double-gate MOSFET, where the channel is made of one of the monolayer 2D materials studied here. The doped contact, oxides (gate and spacer regions) as well as the main device parameters are shown on the figure. The atomic structure that is depicted in this figure is that of a TMD, here it is $HfS_2$, with the metallic atom (Hf) in the center, sandwiched between the 2 chalcogen (S) atoms at the top and bottom.

The 1st step towards transport simulations of a given material is a first-principle geometry relaxation of its primitive unit cell, followed by an electronic-structure calculation. We used the DFT package Quantum ESPRESSO[21] and the generalized gradient approximation with the optB86b exchange-correlation functional.[22] The Bloch wavefunctions are then transformed into maximally-localized Wannier functions (MLWF) typically centered on the ions using the wannier90 package.[23] Figure 2 demonstrates the validity of our MLWF representation for the case of 3 of the 2D materials studied here. The resulting supercell information, including atoms and MLWF positions, lattice vectors, as well as the localized Hamiltonian-matrix elements, are used by ATOMOS as building blocks to create the full-device atomic structure and Hamiltonian matrix. Transport calculations are then performed using a real-space NEGF[24,25] formalism including electron-phonon (e-ph) scattering within the self-consistent Born approximation. [26] More details can be found in the method section.

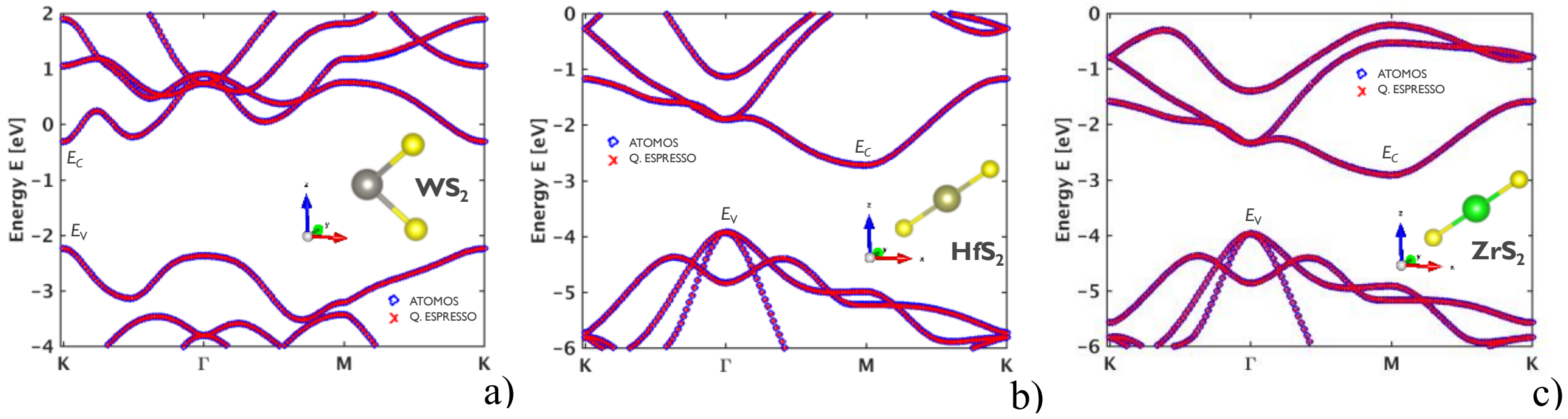

Figure 2. Band structure computed with QUANTUM ESSPRESSO using plane-wave DFT and with ATOMOS using the Wannierized Hamiltonian for a) a monolayer $WS_2$ (2H-phase) b) a monolayer $HfS_2$ (1T-phase) and c) b) a monolayer $ZrS_2$ (1T-phase). The insets also show the atomic structure and chosen cartesian-axes directions for the various supercells.

**2D material screening:**

We first focus on evaluating intrinsic monolayer (1ML) 2D-material device physics and performance using our DFT-NEGF model. The goal is to find a meaningful upper limit to identify the trends and screen the most promising candidates for scaled CMOS applications. Looking at the list of existing TMDs and other 2D materials, we have pre-selected 5 TMDs,[14,16,18] as well as $P_4$,[17,18] due to their relevant electronic and transport properties (band structure, phonon properties, material stability, and/or experimental relevance). For the TMDs, we focus here on $MoS_2$, $WS_2$ and $WSe_2$ in the trigonal prismatic (2H) phase, as these materials are among the most studied and mature experimentally. We also focus on the less explored $HfS_2$ and $ZrS_2$ TMD's (in their most stable octahedral, 1T, phase), as their band structures hint for better transport properties (higher drive current), while retaining a sufficiently high bandgap and balanced properties[16,18] to expect good off-state currents at scaled gate lengths. Finally, current-voltage characteristics for $P_4$ will be investigated here as well, due to the strong attention and expectation this material has stirred in the recent literature.[17]

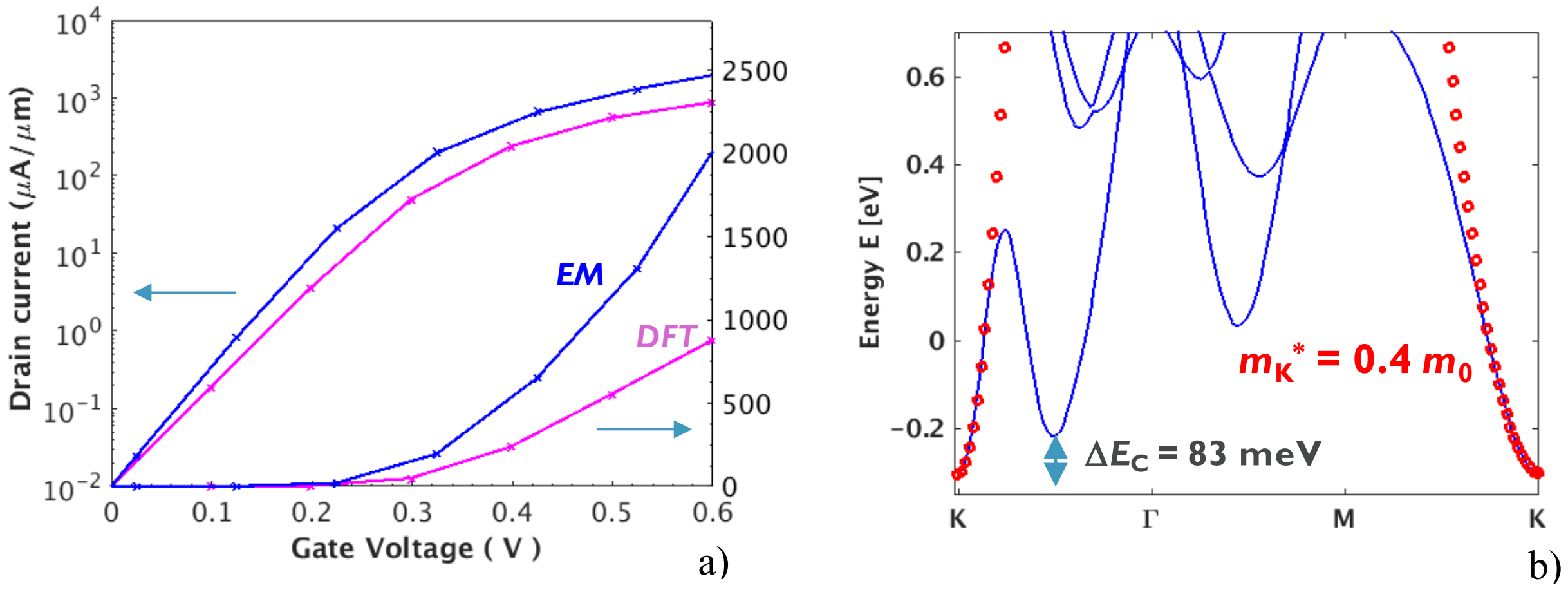

Figure 3. a) Drain-current – gate-voltage, $I_D(V_G)$, characteristics of a $L$ = 5 nm 1ML-$WS_2$-DG nMOSFET computed with DFT and a fitted effective-mass (EM) model (including the 2 first valleys seen in Fig 3b). $V_D$ = 0.6V. b) DFT-computed band structure around $E_C$ (blue line) and that fitted with the 1st valley (centered at $K$-point) isotropic effective mass = 0.4 $m_0$ (red stars), $m_0$ being the free electron mass. Despite that a good agreement between the EM- and the DFT-band-structure model is achieved in the vicinity of $E_C$, the EM-NEGF model strongly overestimates the current drive in the device. The second valley (located between $K$ and Γ) was also fitted with an EM of 0.9 $m_0$ and included in the effective-mass NEGF model.

A) Transport Model Requirement:

Non-atomistic models, such as effective-masses (EM) and derived simplified two-bands NEGF models, although widely used[27,28,29] due to their wider availability and strongly reduced computational cost, are typically inaccurate to model 2D materials. In Fig. 3, it can be seen that although a reasonable fitting of the lowest part of the band structure can be obtained for $WS_2$ using an effective-mass Hamiltonian model (Fig. 3b), the current is strongly over-estimated (Fig. 3a). This is due to the combination of 2 facts. Firstly, due to their specific band structures, e.g., many TMD's (in particular in the conduction band of that made of W or Mo atoms) have narrow energy valleys with discontinuous density-of-state, DoS, profiles that are not captured with simplified band models (e.g., see Fig. 3b, where the 1$^{st}$ $WS_2$ conduction-band valley is a narrow valley with an energy extend that is less than 0.6 eV in the KΓ direction).[30] Secondly, in an extremely thin material, a full atomistic treatment of how the charge is distributed within the 2D layers is required to accurately capture the charge-centroid position. For the case shown in Fig. 3a, as typically the case in TMDs, about 90% of the charge is located on the metallic (W) atom, which is in the middle, not on the surface chalcogenide (S) atoms (Fig. 1). This information is lost in a non-atomistic model, and a homogeneous charge distribution with a centroid closer to the surface is obtained. The effective-mass model is also not able to predict accurately source-to-drain tunneling, hence the subthreshold characteristics of the device (Fig. 3a), a crucial effect for the sub-10 nm gate-length regime where 2D materials are envisioned to be used for CMOS. Finally, using these approximate EM and derived 2-bands NEGF models, it was assessed that a bi-layer DG device could deliver more drive-current than that of a 1ML device for the sub-10 nm HP -CMOS application.[28,29] Our DFT-NEGF results, however, show that a 1 ML material, which is the main focus of this paper, is preferred (more details can be found in the SI section 1). For accurate results, full-band atomistic-transport simulations, such as the DFT-NEGF results presented here, are therefore needed.

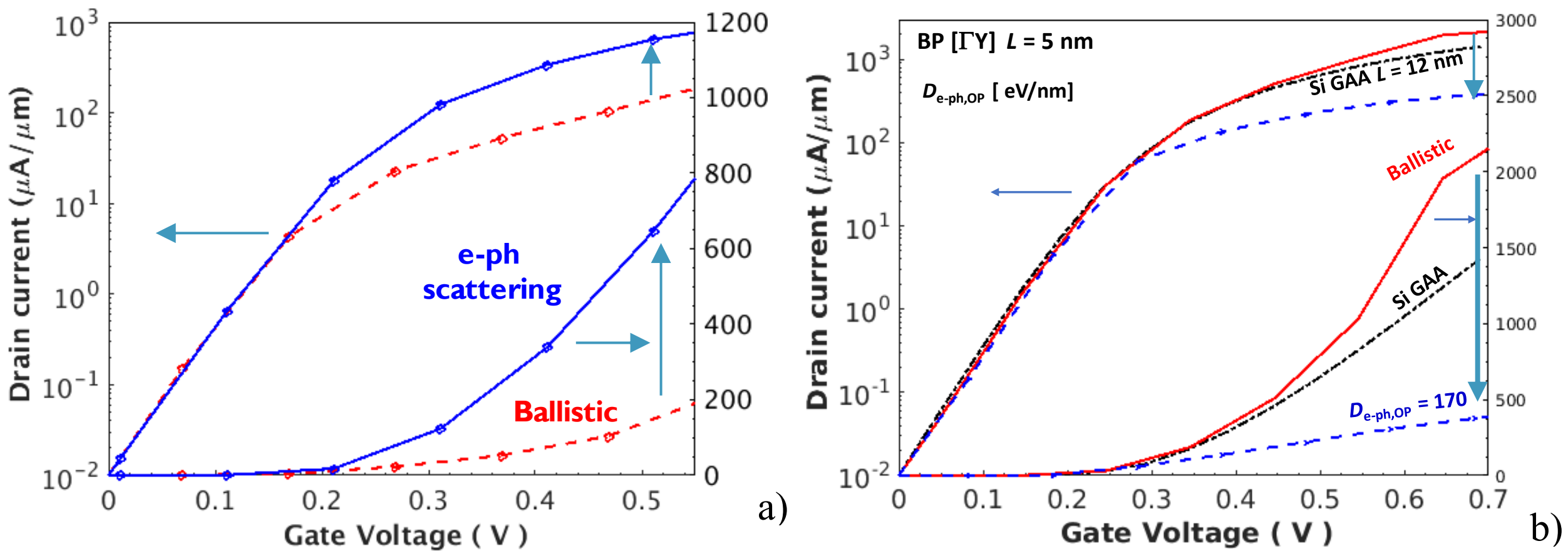

Figure 4. a) $I_D(V_G)$ characteristics of a $L$ = 14 nm bi-layer $WS_2$-DG nMOSFET computed with DFT-NEGF with and without e-ph. $V_D$ = 0.7V. As q.$V_D$ is larger than the width of the bi-layer first valley, the 1st valley electrons cannot travel ballistically from the channel to the drain and the ballistic current is reduced compared to the case with scattering. b) $I_D(V_G)$ characteristics of a $L$ = 5 nm, 1ML-$P_4$-DG nMOSFET computed with DFT-NEGF with and without e-ph. $V_D$ = 0.6 V. Due to the strong optical-phonon coupling ($D_{OP}$ = 170 meV/ nm, $\hbar\omega_0$ = 32 meV),[45] the drive current drops significantly when e-ph is included, despite the very short channel length. $I_{OFF}$ = 10 nA/µm. ΓY (zigzag) channel orientation.

It, however, turns out that for 2D materials, ballistic full-band transport simulations are not accurate enough, even at a gate length as short as 5 nm. The argument often used of very-short $L$ to justify ballistic transport in conventional "3D"-material transistors does not hold true here. In 2D-materials with narrow valleys, a "valley-filtering effects" typically strongly reduce the ballistic current and inelastic-scattering effects need to be included to recover the current in the device[30] as shown in Fig. 4a for a bi-layer $WS_2$ transistor. In other cases, like for black phosphorus, strong optical-phonon modes can significantly reduce the current compared to the ballistic case (Fig. 4b). Thus, a full-band dissipative atomistic treatment, as presented here is needed to get an accurate and meaningful upper limit of 2D-material devices. This upper limit may still be far from today's reality, as we are neglecting interactions with the environment (e.g., contact resistance, surrounding oxides...), and defects that are usually strongly present in nowadays experimental devices.[14,17,20] It, however, gives insight of what is to be the fundamental potential of such a technology, as it matures.[20,31]

B) DFT-computed material parameters and properties:

SI Table 1 (in the section 2 of the supplementary information) gives the relaxed unit-cell dimensions and bandgaps we obtained for the TMDs studied here. They are in good agreement with other DFT calculations in the literature[16,18] and experimental results.[32]

From our NEGF simulations, we have also extracted the electron or hole concentrations *vs.* the Fermi-level, $E_F$, position with regards to conduction- or valence-band edges, $E_C$ or $E_V$, respectively, i.e., $E_F - E_C$ or $E_V - E_F$. By fitting those to an analytical 2D-DoS model, the conduction- or valence-band DoS, $N_{2D}$, as well as an equivalent DoS mass, $m_{DoS}$, can be computed. Both values are reported in SI Table 1 for the TMDs studied here. This $m_{DoS}$ folds the DFT-computed non-parabolicity of the occupied bands close to the conduction- or valence-band edges into a simplified, equivalent, single-band, parabolic effective-mass model (the details are in the method section). SI Figure 4 shows, for a representative sample, the good level of agreement that can be achieved between the analytical charge model and the DFT-NEGF simulated data. $N_{2D}$, or equivalently $m_{DoS}$, as well as the mobilities, that we will extract next, are useful quantities for developing simplified TCAD or compact models and benchmarking 2D-material performance.[33] Note that, in this paper, densities, as well as doping concentrations, are given per unit of volume. Those can be converted to the per unit of area, often used for 2D materials, by multiplying by the 2D-film thickness, $t_S$, about 0.6 nm for a monolayer TMD (the exact value used for each studied monolayer can be found in SI Table 1).

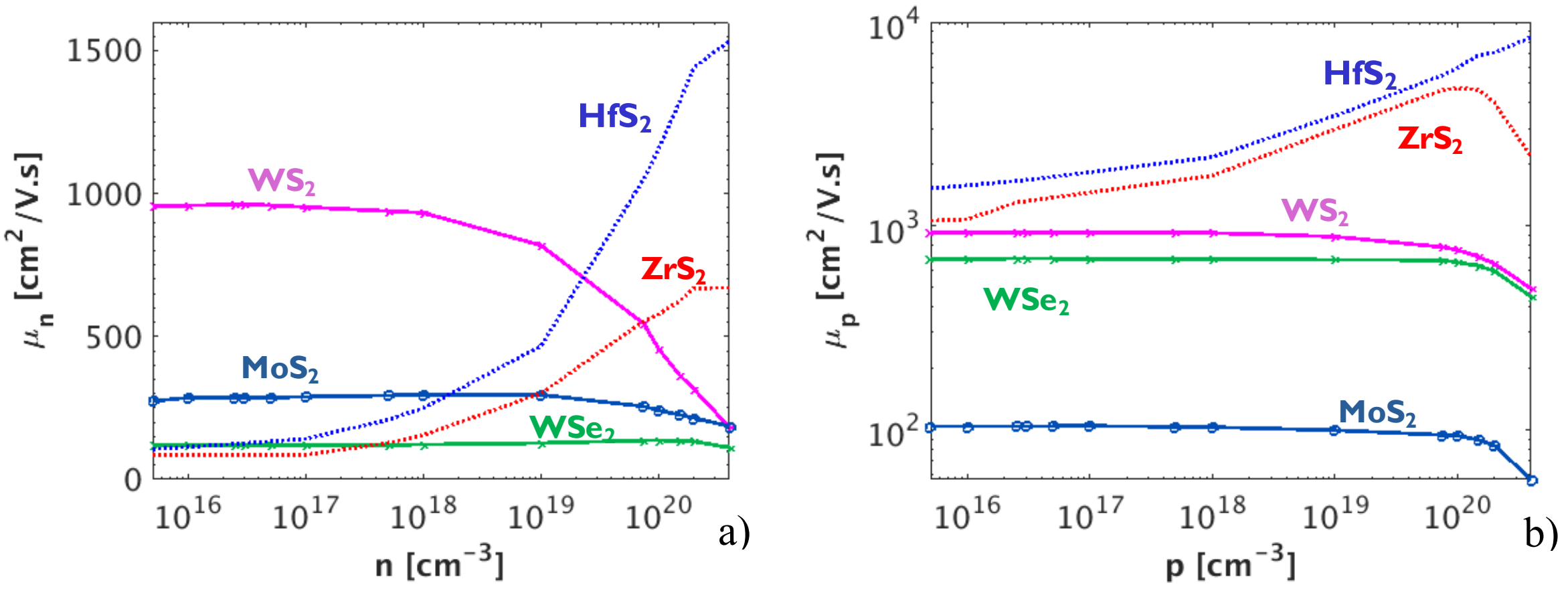

Figure 5. Intrinsic electron-phonon-limited mobility *vs.* carrier concentration in a) n- and b) p-type TMD transistors. The mobilities were extracted from our DFT-NEGF simulations using long-channel devices (for several channel $L$ ranging from 100 nm to 1 µm typically) at $V_D = 50$ mV. The $L$-independent ballistic resistance was removed from the extraction by using the *dR/dL* method.[34] $HfS_2$ and $ZrS_2$ results are shown along the ΓK channel orientation and include polar-optical phonons using a Fröhlich model.

Fig. 5 shows the long-channel low-field intrinsic electron and hole mobilities. Each mobility curve was extracted from 4 to 7 DFT-NEGF simulations using devices with various channel lengths, typically ranging from 100 nm to 1 µm, using the dR/dL method[34] to correct for the ballistic resistance. The intrinsic mobility is a convoluted value, resulting from band structure (intrinsic transport properties) and electron-phonon (e-ph) scattering. It is a key-performance metrics for long-channel devices. For electron-phonons, we used the DFT-computed values of ref.[16] and included the dominant acoustic, and optical modes. For the 1T TMDs that are polar materials, we also include the polar-optical phonons (POP) using a Fröhlich model[35,36] that takes into account the electronic screening and the long-range interaction up to 3 nm (more details can be found in section 3 of the SI). We note that, the long-range LO polar component can be directly included in the DFT calculations, as it was done in.[37] However, a Fröhlich model provides a direct way to consider the screening which is important and was neglected in.[37] 1T TMD material mobilities could also suffer, in principle, from a non-vanishing ZA phonon term due to the lack of horizontal mirror symmetry.[38] In order to verify this, we have computed the $HfS_2$ electron-phonon matrix elements from DFT, using a similar approach that the one used in.[37] We have found that ZA phonons do contribute in $HfS_2$, but that their contribution is small compared to that of the other acoustic phonon modes (TA and LA). A similar conclusion was found for both the mobilities of $HfS_2$ and $ZrS_2$ in. [37]

For the 2H TMDs, the mobility curves typically present a plateau region dominated by intra-band low-energy acoustic eph-scattering. At higher carrier concentrations, when the position of the energy band with regards to $E_F$ is sufficiently degenerated so that satellite energy valleys start to be populated, higher-energy intervalley e-ph scattering mechanisms further degrade the mobilities. The carrier concentration at which this degradation happens depends on the energy separation between the 1st and the satellite valleys and the $m_{DoS}$ (a larger $m_{DoS}$ leads to less degeneracy at a given concentration as illustrated in SI Fig. 4). The NEGF-computed mobility values for the 2H TMDs are in qualitative agreement with mobilities calculated in the literature with various methods,[37,39,40] showing same order of magnitude and ordering. $WS_2$ has the highest mobility. For p devices, $WSe_2$ also features an interesting value.

For the 1T TMDs, the plateau region is not observed in the mobility curves. Their mobility rather increases for increasing carrier concentration. SI Fig. 7 compares the n- and p-type $HfS_2$ total mobilities including POP and screening to that without POP and that with POP but neglecting the screening. The total mobility value is limited by the strong, and nearly unscreened, POP interaction at low carrier concentration. As the carrier concentration increases, however, screening renders POP scattering less efficient and the mobility increases towards the limit without POP. It is to be noted that the n-type mobility value of 1896 $cm^2/V.s$ that we obtain in the plateau region for the case without POP well agrees with the about 1800 $cm^2/V.s$ acoustic phonon-limited value computed in ref. [37] and [39]. The n-type low mobility value of about 60 $cm^2/V.s$ obtained for the case that included POP, but neglecting the screening, is consistent with the results and hypothesis presented in ref. [37].

Finally, as presented in SI Fig. 8 for $HfS_2$, for scaled nanoscale devices, the impact of POP scattering is not significant, owing to the short channel lengths and the strong screening related to the high carrier concentration in on-state. The impact of high-energy optical phonons is typically rather limited in the subthreshold regime of a scaled transistor. Despite the weak screening in the channel, related to the low subthreshold-carrier concentration, most of the electrons are injected at an energy in thermal equilibrium with the top-of-the-channel barrier (e.g., see Fig. 9). The majority of empty states, in which electrons could scatter to, are, however, localized at the same energy. Hence, low-energy acoustic phonons are rather the dominant scattering mechanism in subthreshold regime. In on-regime, the situation is different (e.g., see Fig. 10a) but POP is effectively screened. From the above discussion and results, one concludes that for nanoscale devices with strong polar interactions, the high-density screened mobility is likely to be the relevant one. The rather high mobilities obtained for $HfS_2$ and $ZrS_2$, at high carrier concentration (in on-state carrier densities of several $10^{20}$ $cm^{-3}$ are typical), highlights their interesting transport properties.

By definition, the current is the product of the number of mobile-charge carriers, that is proportional to $m_{DoS}$, times their velocity, that is proportional to the mobility. Hence the $m_{DOS}$ × mobility product of a given material is an indication of its MOSFET drive-current potential. This product, normalized so that it is equal to 1 for the n$WS_2$-case, is reported in SI Table 1 for n- and p-type conduction and allow a relative comparison between the different TMDs reported here. Again, the drive-potential of $HfS_2$ and $ZrS_2$ stands out, while $WS_2$ comes in 3rd position.

C) Sub-10 nm Fundamentals:

2D materials are, however, envisioned to be used in the sub-10nm gate-length regime as potential replacement for Si. At such $L$, the mobility × $m_{DOS}$ product alone is not a sufficient metrics to compare performance. Other effects such as source-to-drain tunneling, that become important due to the narrow channel barrier, typically penalize more high-mobility materials (as the ability of quantum-mechanical tunneling is enhanced in low-effective mass materials) and a trade-off exists. The case of $P_4$ that we will further discuss below is a good example. Similarily, a recent publication has used DFT-NEGF simulations to screen 100 2D materials and found 13 potential candidates with very high drive current potential at $L$ = 15 nm (pending a detailed study on the impact of e-ph scattering).[41] When scaling $L$ down to 5 nm, however, the SS of these devices were all degraded and ranging in the 110 to 275 mV/decade using a DG architecture. It is to be noted that the dynamically-doped-transistor concept, that will be studied in the last part of this manuscript, might be a way to utilize the strong drive potential of such materials at further scaled dimensions.

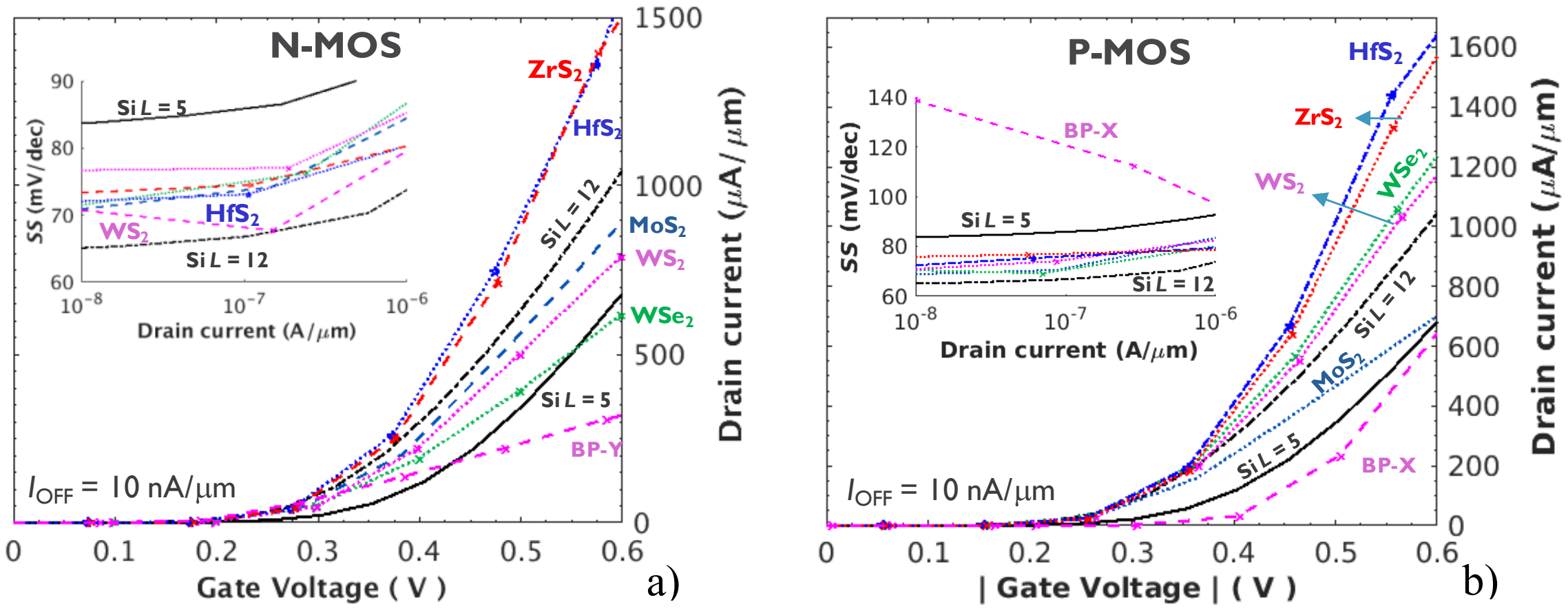

Figure 6. $I_D(V_G)$ characteristics and $SS(V_G)$ (inset) of the optimized $L$ = 5 nm 2D materials DG-MOSFETs, and the $L$ = 5 and 12 nm Si optimized GAA a) n- and b) pMOS transistors. $|V_D|$ = 0.6 V. $I_{OFF}$ = 10 nA/μm. e-ph scattering is included. The devices are optimized in terms of source- and drain-doping concentration ($N_{SD}$), channel orientation (if anisotropic) and thickness ($t_S$) for the GAA. $HfS_2$ and $ZrS_2$ IV's are shown along the ΓK channel orientation. The current is normalized by the gate perimeter.

Fig. 6 compares the $I_D(V_G)$ characteristics of $L$ = 5 nm DG n- and p-MOSFETs made of six different 2D materials (Fig. 1), the 5 TMDs previously studied as well as $P_4$, at a typical HP off-state leakage of $I_{OFF}$ = 10 nA/μm.[12] We further benchmark their performance against that of an optimized Si n-type gate-all-around nanowire (GAA) with a square cross-section, but with a relaxed gate length ($L$ = 12 nm).[42] The optimized $L$ = 5 nm Si GAA $I_D(V_G)$ is shown as well. The GAA were simulated using cleaned mode-space $sp_3d_5s$* tight-binding NEGF models.[36,42]

For Si, scaling $L$ below 10 nm typically results in subthreshold slope ($SS$) and $I_{ON}$ degradation. This is due to electrostatic-control losses, quantum confinement and source-to-drain tunneling. SDT and QC become significant for $L$ < 10 nm and $t_S$ < 4 nm respectively. It was observed that e-ph scattering was significantly enhanced, even at short $L$, for Si GAA with $t_S$ < 4nm,[8,10,26] one reason being the increase of the electron-phonon wave-function overlap in strongly confined wires due to volume inversion.[10] This is one of the fundamental reasons behind the strong mobility reduction observed in thin-film "3D" materials. For the 5nm long Si device, the narrow $t_S$ of 3.5 nm used in the simulations will further result in a strong threshold-voltage variability related to surface-roughness induced bandgap change with QC.[2,10,11]

For all the 2D materials shown on Fig. 6, excepted for the p-type $P_4$-device case that will be discussed below, we observed less $I_{ON}$ and $SS$ degradation than for Si, when scaling $L$ down to 5 nm. This is related to their excellent electrostatic control (a better electrostatic control enables a larger effective channel length at same nominal $L$, hence less SDT) and QC-free characteristics stemming from their 2D-nature.

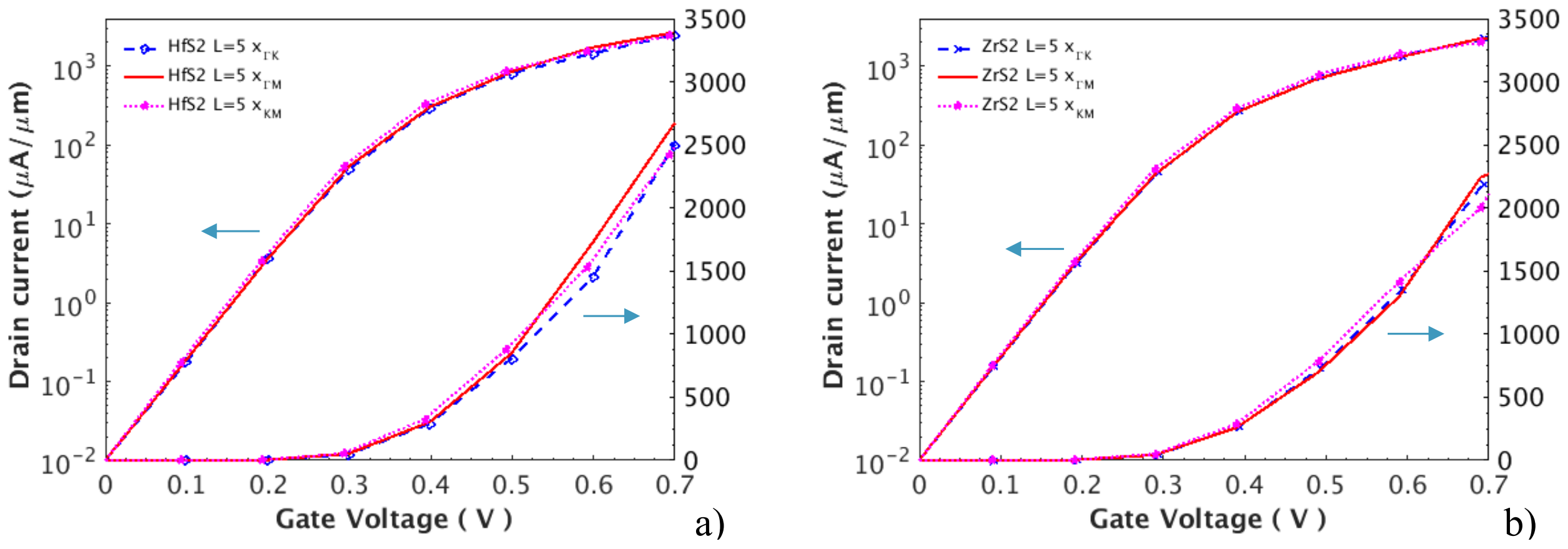

Figure 7. $I_D(V_G)$ characteristics of group-IV TMD nMOSFETs for 3 different channel orientations along principal crystallographic axes (ΓK, ΓM and KM), a) for $HfS_2$ and b) for $ZrS_2$. $V_D$ = 0.6 V. $L$ = 5 nm. $I_{OFF}$ = 10 nA/μm. e-ph scattering is included.

The outstanding performance of the 2 group-IV TMDs, i.e., of $HfS_2$ and $ZrS_2$, that feature, by far, the highest on-current levels is also highlighted on the plot. Besides the afore-mentioned excellent electrostatic control and confinement-free characteristics, this is related to $HfS_2$ and $ZrS_2$ well-balanced transport properties that allow for high $I_{ON}$ with limited SDT. The closely matched characteristics of both $HfS_2$ and $ZrS_2$ can be understood in light of their similar band structure (Fig. 2b and 2c) and the mitigation of the mobility × $m_{DOS}$ product (that is better for $HfS_2$) at very short $L$. It is to be noted that $HfS_2$ and $ZrS_2$ have anisotropic band-structure properties. Fig. 7 shows the impact of crystal orientation on the performance of the $HfS_2$ and $ZrS_2$ n-type devices. Their performance is not varying much along the principal-axis directions shown here. Overall, group-IV TMDs show promise for scaled HP CMOS.

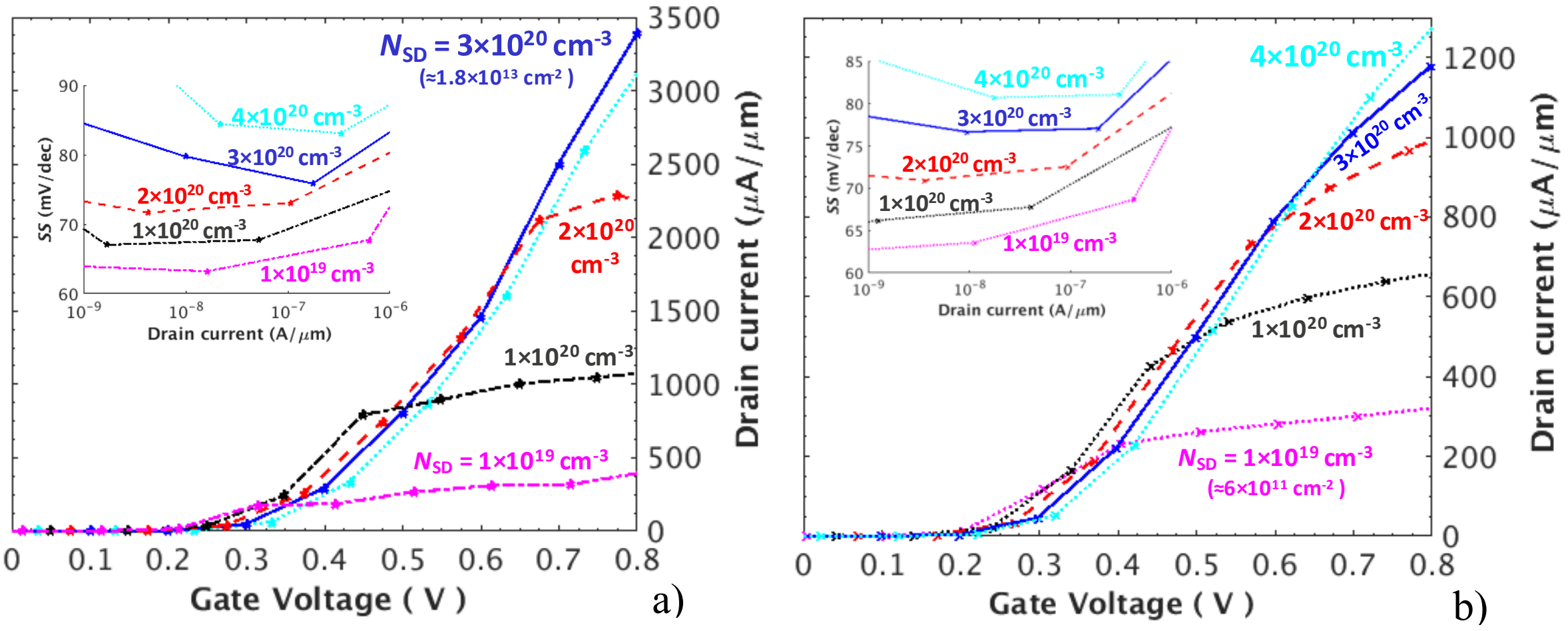

Figure 8. $I_D(V_G)$ characteristics and $SS(V_G)$ (inset) of 2D nMOSFETs vs. $N_{SD}$ a) for $HfS_2$ and b) for $WS_2$. $V_D$ = 0.6 V. $L$ = 5 nm. $I_{OFF}$ = 10 nA/µm. e-ph scattering is included. The trade-off between on-state (better for higher $N_{SD}$ values, due to a reduction of source starvation) vs. off-state (better for lower $N_{SD}$ values, due to a reduction of SDT) is observed for both materials.

From the more studied group-VI TMD family, $WS_2$ emerges as the best candidate for n and p on average, i.e., second best and close to $MoS_2$ for n, and best with $WSe_2$ for p. $MoS_2$ performs poorly both for p-type, while $WSe_2$ performs poorly for n-type conduction. This can be correlated to the intrinsic transport properties (e.g., see mobility×$m_{DoS}$ in Table 1) of these materials for p-type. For n-type an additional factor has to be considered. Excepted for $MoS_2$ that has very poor p-type performance, group-VI TMDs have a markedly stronger p-type drive-current than that of the n-type. This is related, at least to a great extent, to the narrow valleys that are present in the conduction band of the Mo- or W-based TMDs, as discussed above. These prevent, at least partially, direct ballistic current from the channel to the drain at $V_{DS} \geq$ 0.6V, so that a less-efficient phonon-assisted transport is required. 1ML-$WS_2$ or $MoS_2$ feature a relatively wide 1st conduction-band valley (for $WS_2$, for instance, its width is actually not isotropic and can be especially large in certain orientation such as the KM orientation shown on Fig. 2a and 3b). The combination of this fact with its higher electron mobility×$m_{DoS}$ product explains $WS_2$ good position for n-type transport in this group.

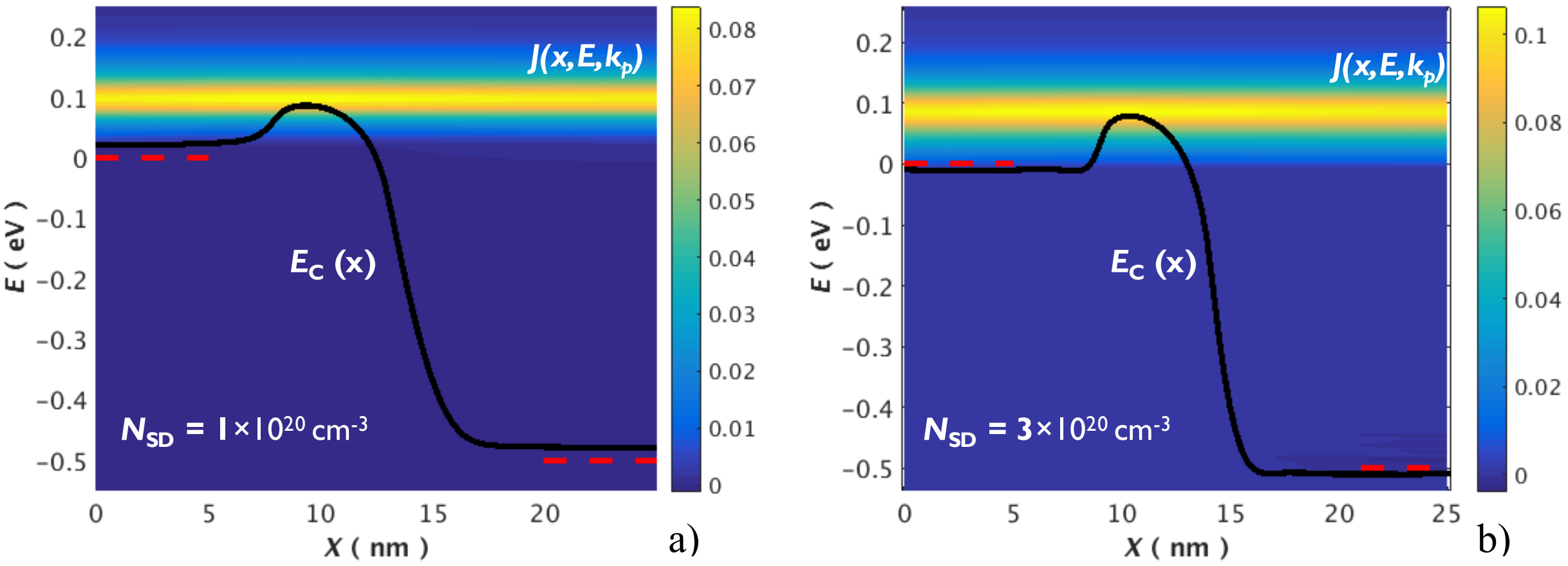

Figure 9. Current spectrum *J(E, x)* (surface plot), as well as top conduction-band ($E_C$ ) (-) edge along the channel direction, $x$, of a $HfS_2$ nMOS with $L$ = 5 nm, in off-state. a) For $N_{SD}$ = 1× $10^{20}$ cm$^{-3}$. In this case, the lower $N_{SD}$ value is not sufficient to ensure degeneracy of the extension as can be seen at the source and drain-sides, where the conduction band is above the Fermi-levels, $E_{FS}$ and $E_{FD}$ respectively. Both Fermi levels are indicated by a red dashed line. b) For $N_{SD}$ = 3× $10^{20}$ cm$^{-3}$. The resulting narrower channel barrier allows for a larger part of the current spectrum to tunnel under the channel barrier (SDT). This effect is enhanced for larger $N_{SD}$ values as the effective channel length is reduced. $V_D$ = 0.5 V.

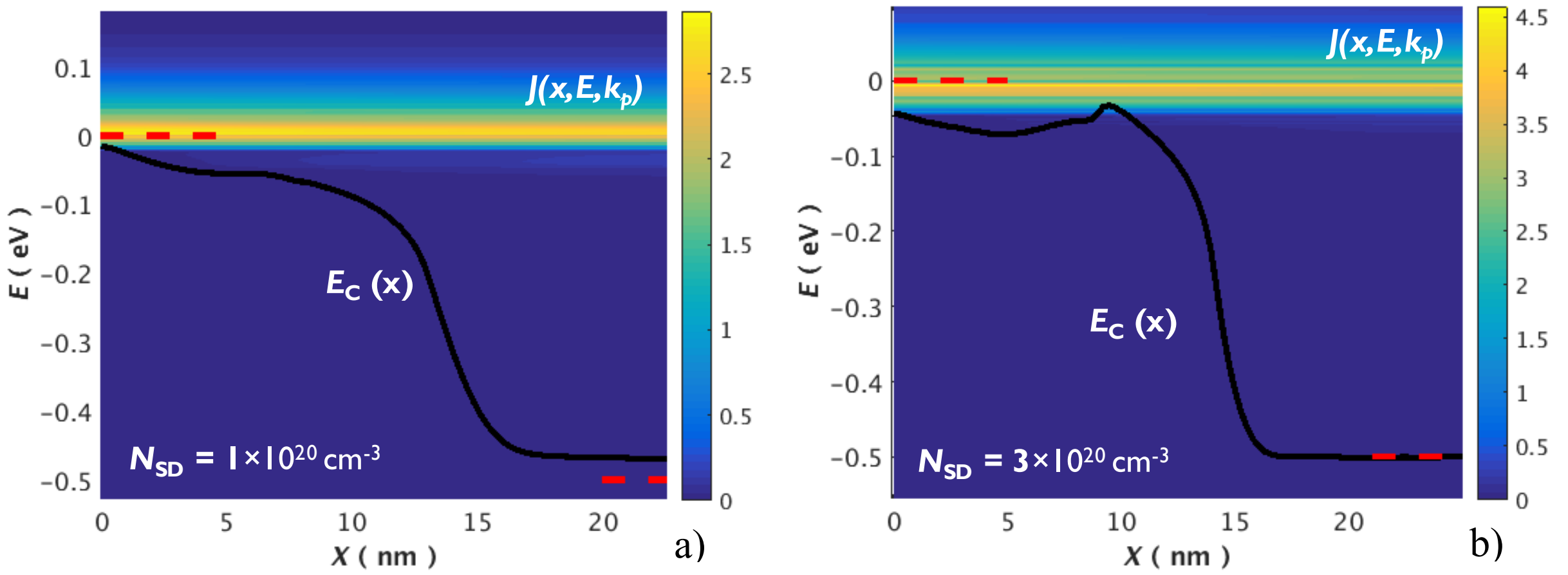

Figure 10. Current spectrum *J(E, x)* (surface plot), as well as top conduction-band ($E_C$ ) (-) edge along the channel direction, $x$, of the $L$ = 5 nm $HfS_2$ nMOS in on-state at $V_G$ = 0.6V. a) For $N_{SD}$

= 1× $10^{20}$ $cm^{-3}$. In this case, due to the lower $N_{SD}$ value, the current flow is limited by the source conduction band and source starvation is observed in the $I_D(V_G)$ characteristics. b) For $N_{SD}$ = 3× $10^{20}$ $cm^{-3}$. In this case, the current is still limited by the channel barrier that is well controlled by the gate. Current saturation is not yet observed in the characteristics at this gate voltage. $V_D$ = 0.5 V.

The $P_4$ device shows the worst performance, which might come as a surprise. P4 is a strongly anisotropic material with a low effective mas in the ΓX direction (i.e., high mobility) and a high effective mass in the ΓY direction (i.e., low mobility). This is true both for n- and p-type conduction (more information can be found in section 4 of the SI). Hence, this material has a very strong drive current for longer channel in the ballistic regime and the ΓX transport direction. In the sub-10 nm regime, however, the armchair (ΓX) oriented $P_4$ transistor, strongly suffers from SDT and is hard to switch-off due to its very-low transport effective mass. For $L$ < 10 nm, the strongly anisotropic $P_4$ material was shown to perform best in the zigzag (ΓY) orientation for the n-type transistor.[19] Still, the ΓY $P_4$ n-device shows a strong drive current in the ballistic regime (Fig. 4b). Most theoretical studies on scaled $P_4$ devices have looked either at ballistic performance and simplified band models,[43] or have neglected the optical-phonon coupling.[44] The ΓY $P_4$ nMOSFET $I_{ON}$ is strongly degraded by its optical-phonon (OP) coupling ($D_{e\text{-}ph,OP}$ = 170 eV/nm for a single monolayer)[19,45] (Fig. 4b). Concerning the pMOS, the ΓX transistor still performs the best at $L$ = 5 nm but its drive-current is severely degraded by SDT (Fig. 6b). The ΓY drive-current is indeed even lower than in the n-case, while the SDT-related sub-10 nm SS degradation in the ΓX direction is not as strong as for the n-case (more details are available in the SI section 4).

For all the 2D materials studied here, we found that a high doping density in the source and drain extensions, $N_{SD}$, ($N_{SD}$ = 2 to 4 × $10^{20}$ $cm^{-3}$) had to be used to reach their fundamental current level. Fig. 8. shows the impact of $N_{SD}$ on the current for the $L$ = 5 nm $HfS_2$ and $WS_2$ nMOS devices. Fig. 9 and 10 give details on the current spectrum flow and the conduction band position in the $HfS_2$ transistor for $N_{SD}$ = 1 and 3 × $10^{20}$ $cm^{-3}$ in off- and on-state respectively. Due to the typically high density of states in these materials (about 30× (10×) that of Si for $HfS_2$ ($WS_2$ respectively)) (see SI Table 1), a high $N_{SD}$ value is required to fully degenerate S&D extensions (Fig. 9) and avoid a saturation in the $I_D(V_G)$ characteristics in on-regime related to a source-starvation effect.[46] In the case of source starvation, as the gate voltage is increased in on-regime, the current is limited by the availability of carrier in the source. On Fig. 10a, it can be seen that, when the source-starvation regime is reached, the current is not limited by the channel barrier. It is rather limited by the energy band at source side. The latter is only indirectly and weakly affected, when switching on $V_G$, by the increase of the non-equilibrium transport charge through the device. This leads to a weak increase and eventually a saturation of the current in the $I_D(V_G)$ characteristics. By increasing the source- and drain-extension doping to ensure good degeneracy at the source side, however, this effect is delayed to higher gate-overdrive values (Fig. 10b and 8). As can also be seen on Fig. 8, increasing $N_{SD}$ has a detrimental effect on *SS* at such a scaled gate length, so that an optimal value exists. This is related to a reduction of the effective-channel length and an increase of SDT for higher $N_{SD}$ values (Fig. 9). A similar trend is observed for all the n- and p-type devices studied here and an optimal value between $N_{SD}$ = 2 to 4 × $10^{20}$ $cm^{-3}$ is observed for $L$ = 5 nm and $V_{DD}$ = 0.6 V in all cases.

**The Dynamically-Doped Field-Effect Transistor:**

The trade-off between on- and off-state for the optimal doping concentration becomes more stringent as $L$ is reduced, ultimately degrading transistor performance and preventing further downscaling. Even using 2D materials, scaling below 5-nm gate length becomes very challenging. The case of the monolayer $HfS_2$ transistor with $L$ = 3 nm is shown on Fig. 11a. Using $N_{SD} = 2\times 10^{20}$ cm$^{-3}$ results in strongly degraded SS due to short-channel effects and SDT. The optimal $N_{SD} = 1\times 10^{20}$ cm$^{-3}$ value, however, has poor performance. It suffers both from source-starvation-related on-current saturation and degraded slope due to SDT and SCE.

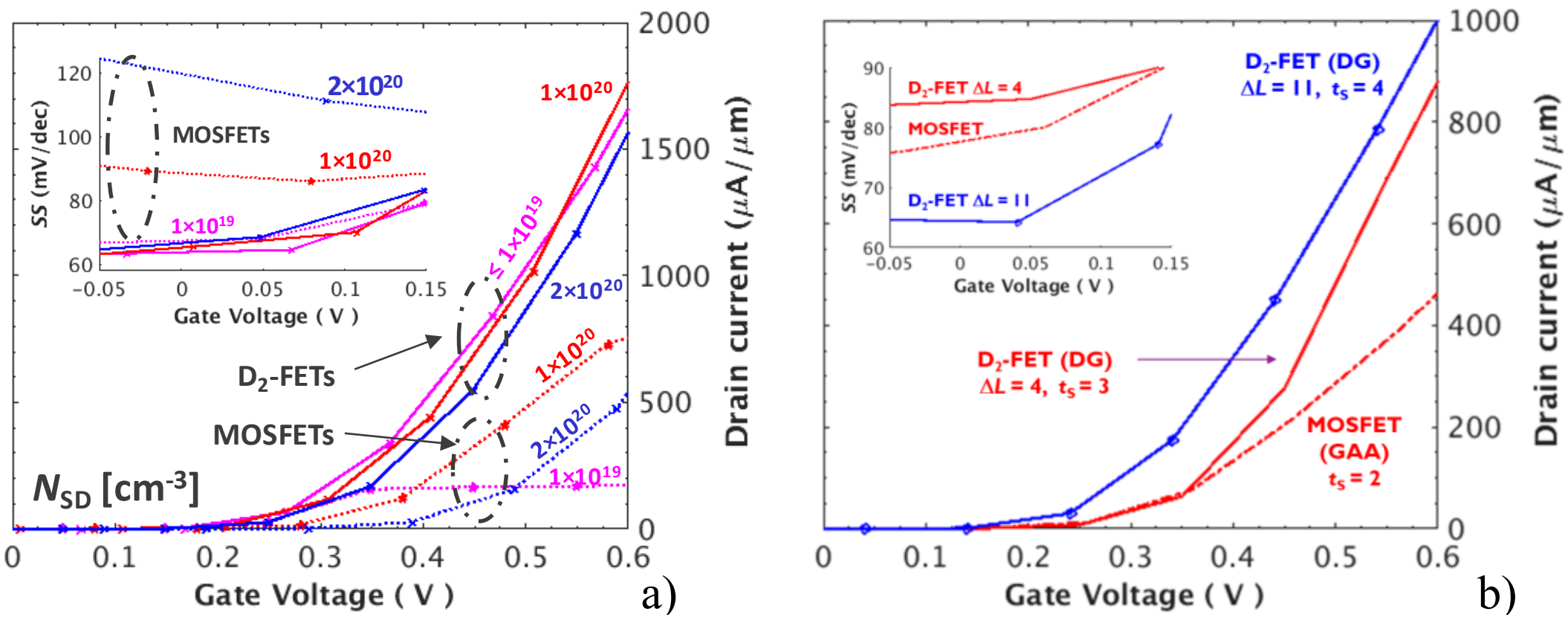

Figure 11. $I_D(V_G)$ characteristics and $SS(V_G)$ (inset) for a) $HfS_2$ double gated nMOSFETs and $nD_2$-FETs with $\Delta L$ = 4 nm *vs.* $N_{SD}$, and b) optimized Si GAA nMOSFETs ($N_{SD} = 1\times 10^{20}$ cm$^{-3}$) and DG $nD_2$-FETs (intrinsic $N_{SD}$) for $\Delta L$ = 4 and 11 nm. For the $D_2$-FETs the optimized film thickness $t_S$ (also indicated in the figure in nm) is larger than for the MOSFET. $V_D$ = 0.6 V. $L$ = 3 nm. $I_{OFF}$ = 10 nA/µm.

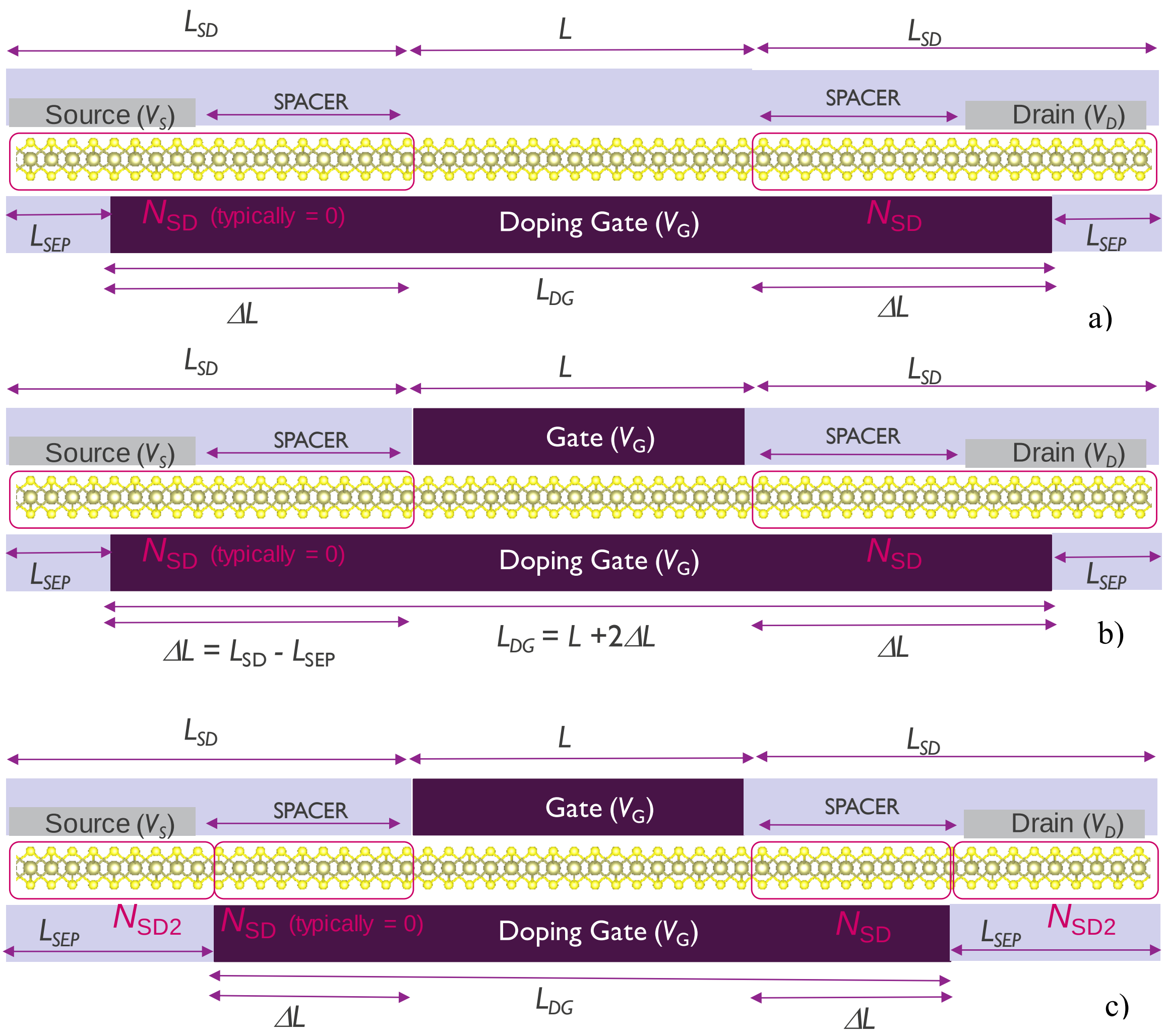

Figure 12. Schematic view of a 1ML 2D $D_2$-FET with a) a single gate, b) a double gate and c) a alternative double-gate design with chemical doping in the source-and-drain-contact regions ($N_{SD2}$), larger $L_{SEP}$ and shorter $\Delta L$.

In addition, as transistor dimensions are scaled down, it becomes increasingly difficult to dope, activate and control the location of the source-and-drain-extension doping atoms, using traditional implantation and annealing techniques.[4,12,20,47,48] In thin-film technologies, direct implantation usually results in a high defect concentration. It is especially the case for 2D materials, where finding a proper way for doping is still an active topic of research.[12,20,49] Typically, more complex techniques, such as epitaxial regrowth of the source and drain extensions, using in-situ doping during the epitaxy, are needed.[12,20,50] Strict control and positioning of the high-doping concentration to prevent its unwanted diffusion during the high-temperature steps of the fabrication process (e.g., during dopants activation or epitaxial-growth phases) in the channel is also challenging and requires advanced techniques such as Flash and Laser anneal,[4,12,47] a solution to this particular problem is to use a uniformly doped, or junctionless transistor.[4] In any case, the discrete nature and limited numbers of doping atoms, resulting from the scaling of the device dimensions, leads to a strong and unavoidable statistical variability when using doping impurities at very small dimensions.[48]

Due to the challenge of chemical doping, electrostatic doping is sometime used in today's experimental 2D or carbon-nanotube material devices to dope their source and drain extensions.[51] It consists in using a gate, e.g., the wafer back gate, to electrostatically induce a high carrier concentration and decrease the semiconductor resistivity, which is the desired effect of chemical doping. Using electrostatic doping with a gate is indeed free of all the afore-mentioned problems related to chemical doping. It can "dope" (i.e., control the carrier concentration in) the entire thickness of the semiconductor film as long as this film is sufficiently thin, typically for $t_S < 10$ nm ( the exact value also depends on the residual chemical doping of the film and substantially decrease if this residual doping is larger than $1\times10^{20}$ $cm^{-3}$). Directly, using the wafer back gate is, however, not a manufacturable solution to individually control billions of transistors with different n- and p-type doping on a chip. A local, dedicated gate for each transistor would rather be required for that purpose. Typically, also these techniques are meant to induce a fixed amount of doping in the source and drain extensions of the device, while the dynamic-doping implications of this technique have not yet been studied. It would indeed be advantageous to have no or a low carrier concentration in the off-state to minimize leakage and a high carrier concentration in the on-state to maximize drive current, i.e., we want to dynamically control doping with the gate voltage of the transistor to break free of the $N_{SD}$ optimization trade-off.

We propose here a novel dynamically-doped FET, which purpose is to turn the challenge of scaling into an opportunity (thin-film and multi-gate architecture technologies, that are the by-product of scaling, enable the manufacturability of such a device). It consists of a transistor that is dynamically doped by one of its own gate(s). This doping gate is located opposite (e.g., at the bottom) the source and drain metal contacts (e.g., located on top). Due to its opposite position, the doping gate, unlike a conventional gate of length $L$, can now overlap, by a value $\Delta L$ on each side, the source and drain extensions to dynamically induces doping without increasing the footprint of the transistor (Fig. 12). This unconventional gate-positioning scheme would alleviate the need for strict alignment control between the doping-gate and the other gates in a multi-gate technology. We insist here that the $D_2$-FET remains a 3-terminal device, like a conventional MOSFET. The doping gate is also the gate of the device for a single-gate device and share the same contact-voltage bias that any other conventional gates, if a multi-gate architecture is used (Fig. 12). It, therefore, does not require an additional contact compared to a MOSFET.

It should be reminded here that, scaling $L$ is a way to scale the total contacted gate pitch, CGP, of a transistor, i.e. the minimal distance between the gate of 2 subsequent transistors. CGP is composed of the sum of $L$ and the length of the highly-doped source-and-drain extensions, $L_{SD}$ (see Fig. 1 and 12). $L_{SD}$ is the sum of the spacer length, that separate the gate contact from the source and drain metal contact of the 3-terminal transistor, as well as the metal-contact length.[12] The length of the doping gate is $L_{DG} = L + 2^* \Delta L$. Technological requirements impose that $\Delta L = L_{SD} - L_{SEP}$, $L_{SEP}$ being a separation distance (typically at least half the spacer length, i.e. a few nm) [12] needed to separate the doping gate from one transistor to that of the next. It is therefore longer than $L$, although it does not require a larger CGP footprint than a traditional top-sided gate of length $L$. To quantify this, the 2031 IRDS-dimensional targets for the so-called 1-nm-technology node and beyond are $L$ = 12 nm, $L_{SD}$ = 14 nm and CGP = 40 nm.[12] The spacer length is 6 nm so that $L_{SEP} \geq$ 3nm, $\Delta L \leq$ 11 nm and $L_{DG} \leq L$ + 22 nm. As can be seen, a comfortable margin is available for $L_{DG}$. This allows for keeping a good electrostatic control, as well as a relaxed $t_S$ scaling, as can be seen for the Si case in Fig. 12b, even when using a very aggressive pitch scaling ($L$ = 3 nm).

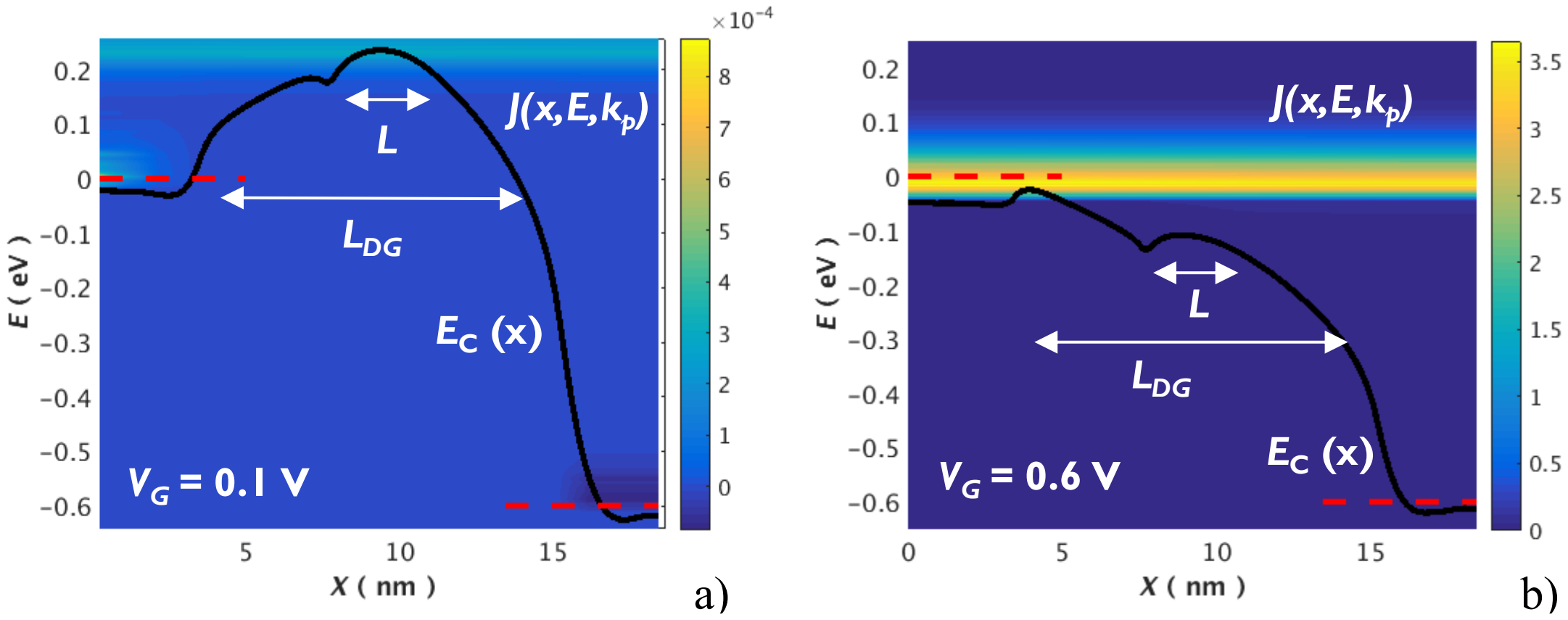

Figure 13. Current spectrum $J(E, x)$ (surface plot), as well as top conduction-band ($E_C$ ) (-) edge along the channel direction, $x$, of the simulated intrinsic ($N_{SD} \leq 1\times 10^{19}$ cm$^{-3}$) $L$ = 3 nm $HfS_2$ $nD_2$-FET of Fig. 11a (with the design of Fig. 12c) a) in off-state at $V_G$ = 0.1V, and b) in on-state at $V_G$ = 0.6V. $\Delta L$ = 4 nm. $N_{SD2} = 4\times 10^{20}$ cm$^{-3}$.

In the rest of this section, if not specified otherwise, we have, however, assumed a worst-case scenario for the $D_2$-FET with smaller $L_{DG}$. We used $\Delta L = L/2$ with a minimum value of $\Delta L$ = 4 nm for $L \leq$ 8 nm. This either assumes a very aggressive $L_{SD}$ scaling, or the possibility of choosing a smaller $\Delta L$ value (i.e., $L_{SEP} >>$ 3 nm) in conjunction with high-doping, $N_{SD2}$, in the $L_{SEP}$ ungated part of the device (as shown in Fig. 12c) to further reduce the contact resistance or for reducing the intrinsic gate capacitance (see discussion below) for instance. Our simulation results show that both cases of $D_2$-FETs (Fig. 12b and 12c) achieve similar $I_{ON}$ and SS for same $L$ and $\Delta L$ in case of ohmic or low Schottky-barrier contacts. The second scheme (Fig. 12c) could be advantageous in case of a high Schottky-barrier contact.

On Fig. 11a, several $L$ = 3 nm DG $HFS_2$ $D_2$-FET characteristics are shown, one with no intentional doping ($N_{SD} \leq 1\times 10^{19}$ $cm^{-3}$, that correspond to typical residual doping concentrations in the 2D films, i.e., lowly-doped or "intrinsic" extensions) and 2 with highly-doped extensions ($N_{SD}$ = 1 and $2\times 10^{20}$ $cm^{-3}$). Our simulations show that for $N_{SD} \leq 1\times 10^{19}$ $cm^{-3}$, the presence of a residual doping in the extensions has no impact on the current-voltage characteristics. The carrier concentration in the extensions is mainly determined by the doping-gate bias. In off-state, the conjunction of a low carrier concentration in the extensions and the extended doping-gate geometry allows for a large $L_{eff}$ (typically $\geq 2\times L$, Fig. 13a) and nearly ideal SS and low off-state current is achieved. In on-state, a high carrier concentration allows for a high drive current. As can be seen, the intrinsic DG-$D_2$-FET, free from any chemical doping, already strongly outperforms the optimized $N_{SD} = 1\times 10^{20}$ $cm^{-3}$ DG-MOSFET. On Fig. 13b, however, it can be seen that in on-state, for a large gate overdrive, the current might be limited by the source part of the conduction-band barrier that is mostly controlled by the doping-gate, not by the top-gate.

In case a large additional $N_{SD}$ doping is used as an attempt to further boost the on-state current, the carrier concentration in the extensions is still dynamically controlled by the doping gate, but the "dynamic-doping" level at a given $V_G$ can be enhanced *vs*. the intrinsic case. In Fig. 11a, it is observed that the current drive can be slightly increased for $N_{SD} \geq 1\times 10^{20}$ $cm^{-3}$, due to the enhanced carrier concentration in the source, while SS is only slightly affected as the doping gate still deplete the extension in the off-state. As in the case of the regular MOSFET, an optimal doping of $N_{SD} = 1 \times 10^{20}$ $cm^{-3}$ is observed for $V_{DD}$ = 0.6V. On the contrary to the MOSFET case, however, $I_{ON}$ and SS sensitivity to doping variations are strongly reduced, and $I_{ON}$ remains high, while SS remains low for all the simulated $N_{SD}$ values. Finally, SI-Fig. 10 compares the $I_D(V_G)$ characteristics of SG and DG 1ML-$HfS_2$ MOSFETS and $D_2$-FETs for $L$ = 3 nm and for $L$ = 5 nm. It is shown that for the $D_2$-FET case, a simpler-to-fabricate SG architecture is as good or even better in term of drive current than that of a DG- $D_2$-FET. The SG-$D_2$-FET indeed keeps a similar and good electrostatic control (SS), when compared to that of the DG- $D_2$-FET case, hence similar drive-current (per gate). For $L \leq 3$ nm, the SG-$D_2$-FET $I_{ON}$ typically outperforms that of the DG- $D_2$-FET device as the short top gate drive only a small amount of additional current compared to the doping gate of the device. In the MOSFET case, a SG architecture is not sufficient to maintain a good electrostatic control for sub-10 nm devices. The SG device SS and $I_{ON}$ is hence degraded compared to the DG-MOSFET case.

Fig. 11b compare $L$ = 3 nm optimized MOSFETS and $D_2$-FETs for the Si case. For the Si-$D_2$-FETs, the number of gates can be reduced, similarily to what was found for the $HfS_2$ case, and an intrinsic DG device was used instead of a GAA. Furthermore,the $t_S$ scaling was relaxed towards $L$ rather than ½ $L$, assuming $\Delta L$ = 4 nm. This strongly reduces QC and boost $I_{ON}$ of the $D_2$-FET, as for the square cross-section GAA the confinement is both in the width (y-) and height (z-direction). In the rectangular cross-section DG-case, the width (y-direction) is typically large compare to its height $t_S$ that is further relaxed compared to the GAA case. In case $\Delta L$ = 11 nm is used, the optimal $t_S$ for the $D_2$-FET is even further relaxed to 4 nm and the performance are further boosted as SS is strongly improved. For Si, we found that the intrinsic case (i.e., unintentionally doped extensions with $N_{SD} \leq 1\times 10^{19}$ $cm^{-3}$) is always better than the case with a larger $N_{SD}$. In any case, even in case chemical doping would be used in the $D_2$-FET, the related challenges (e.g., variability) would be reduced, one reason being the relaxed dimensions ($L_{DG}$, $t_S$) at same $C$GP.

**Scaling Perspective:**

Next, we investigate in Fig. 14 the scaling behavior for both optimized MOSFETs and $D_2$-FETs made of the less explored $HfS_2$, the material showing the highest mobility and performing the best at $L$ = 5 nm, and that of $WS_2$, the best performing material of the more studied group-VI TMDs. The optimized device results for the more conventional Si technology are shown as well. The evolution of their on-current vs. $L$ at a fixed off-state current of 10 nA/μm and a supply-voltage $V_{DD}$ = 0.6V is shown on the figure.

For the MOSFET case, all materials show degradation of their performance when $L$ is scaled below 7.5 nm. This is related to short-channel effects and SDT as discussed before. Comparing the Si-GAA and $WS_2$ DG MOSFETs that achieve similar drive current at $L$ = 10 nm, it can be seen that the GAA $I_{ON}$ degrades faster when downscaling $L$. This is related to the better electrostatic control and QC-free characteristics of the 2D material over the GAA. None of these 2 materials, however, are able to meet the stringent high-performance IRDS $I_{ON}$ targeted for year 2031[12] and a higher-mobility channel material is needed. $HfS_2$ on the other hand, owing to its outstanding transport properties, exceeds the target down to a channel length of about 6 nm.

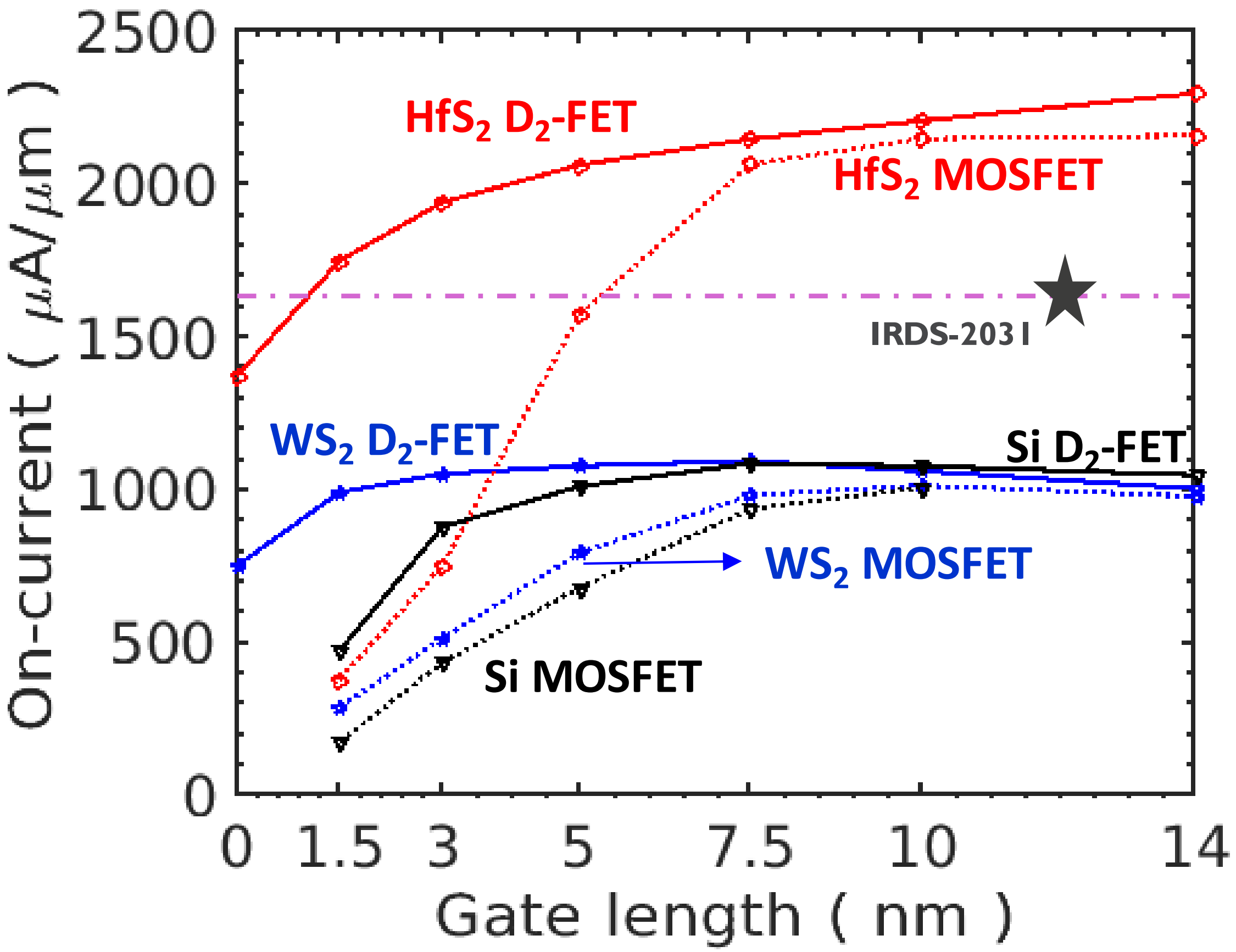

Figure 14. Maximum achievable $I_{ON}$ vs. $L$ for optimized n-type MOSFETs and $D_2$-FETs made of Si, $WS_2$ and $HfS_2$. A DG architecture is assumed for all the 2D-MOSFETs. For the Si MOSFETs, a GAA transistor was used. A SG-architecture is employed for the 2D $D_2$-FETs, while a DG transistor is used for the Si $D_2$-FETs. EOT = 0.5 nm. $V_{DD}$ = 0.6 V. $I_{OFF}$ = 10 nA/μm. The current is normalized by the gate perimeter. For the $D_2$-FETs, we used $\Delta L = L/2$ with a minimum value of $\Delta L$ = 4 nm for $L \leq 8$ nm.

Using the $D_2$-FET concept, it can be seen that the performance degradation with $L$ is delayed to $L$ below 5 nm in all cases. For such small gate lengths, the drive-current potential is strongly enhanced compared to that of the MOSFET case. At $L$ = 3 nm, the Si, $WS_2$ and $HfS_2$ $D_2$-FETs show a $I_{ON}$ gain of about 3× compared to their MOSFET counterparts. It is now possible to reach the IRDS target with a $L$ of about 1 nm $HfS_2$ SG $D_2$-FET transistor (using $\Delta L$ = 4 nm).

Our simulations finally show that, using this scheme, it is still possible to have a transistor effect using $L$ = 0 nm, and that this device performs as well as a regular chemically doped multi-gate transistor with a 4nm or longer $L$ values for the case with 2D materials, although using only single-gated intrinsic semiconductor materials. This shows the promise of using the $D_2$-FET concept for sub-10 nm transistors and in particular for ultra-scaled high-mobility material devices that would be required to meet the stringent IRDS HP targets.

It is to be noted that the larger $L_{DG}$, will however increase the intrinsic gate capacitance at same $L$ and a trade-off between $\Delta L$ and the $I_{ON}$ gain may exist for the speed performance of the $D_2$-FET. In modern scaled technologies, the total load capacitance of inverters or other digital circuits is often dominated by extrinsic (back-end-of-line) capacitances, such as the metal-line capacitance that is proportional to CGP (not $L_{DG}$).[12] By enabling further downscaling with strongly improved drive-current, the $D_2$-FET may therefore also enable a power-delay benefit. To investigate the trends, we performed, here, a power-delay performance analysis of the basic building block of a digital circuit, i.e., a CMOS inverter, using scaled $D_2$-FET and MOSFET devices. The details about the process assumptions and layout, that determine the number of stacked-transistors per device and their geometries, as well as their back-end-of-line capacitance load that is considered in this analysis are detailed in the text of SI section 5 and SI Fig. 11 and 12.

Fig. 15 compares the switching energy *vs.* delay (EDP) of back-end-loaded high-performance invertor cells made with $HfS_2$ DG MOSFETs and $HfS_2$ SG $D_2$-FETs, as well as that made with Si GAA's and Si DG $D_2$-FETs (typically the best device architectures per category of materials for MOSFETs and $D_2$-FETs). More detailed analyses to identify the best devices per category is available in SI Fig. 13 and surrounding text. The inverters are loaded with a typical 50 contacted-gate pitch-long metal line.[12] As CGP is reduced for shorter $L$, more aggressively scaled devices get a net capacitance reduction. The SG-$D_2$-FET devices that only require 1 spacer length (hence a reduced CGP) instead of 2 (see SI Fig. 11) get a further reduction at same $L$. The extrinsic capacitances of the cell layout are also included in the load capacitance. Again, the SG-$D_2$-FET devices are free from the extrinsic parasitic components $C_{GSext}$ and $C_{GDext}$ (see SI Fig. 10 a and b) as the gate metal contact does not directly face the source and drain metal contacts.

In Figure 15, $V_{DD}$ is varied from 0.4 to 0.7 V. The best EDP performance is achieved by the $L$ = 0 nm, $HfS_2$ SG $D_2$-FET, that uses the simplest (SG) architecture and further yields the largest pitch reduction (CGP = 22 nm). It is closely followed by the $L$ = 5 nm, $HfS_2$ DG MOSFET (CGP = 33 nm) and the $L$ = 3 nm, $HfS_2$ SG $D_2$-FET (CGP = 25 nm). The $L$ = 3 nm, $HfS_2$ DG MOSFET (CGP = 31 nm) performance is strongly degraded, and about equivalent or worst (the speed performance at high $V_{DD}$ saturates due to on-current saturation) to that of the $L$ = 5 nm Si DG $D_2$-FET (CGP = 33 nm). The latter 2 devices still comfortably outperform the $L$ = 12 nm Si GAA (CGP = 40 nm).

These results further confirm and highlight the promising potential of the D2-FET device, paving the way towards ultimately scaled devices, with reduced process complexity and variability (e.g., reduced number of gates, larger $t_S$, doping free or reduced sensitivity to doping fluctuations...) and improved performance.

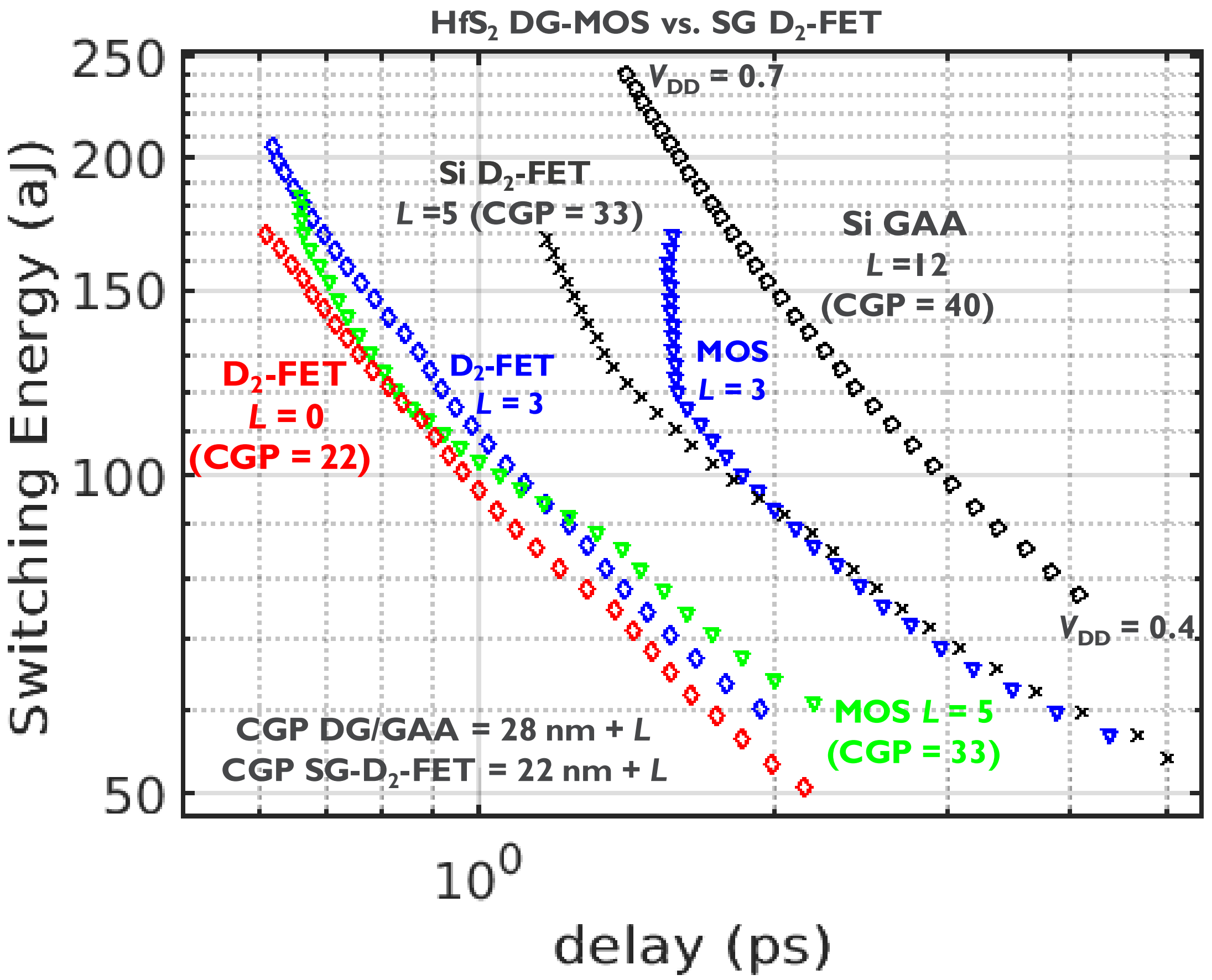

Figure 15. Switching energy *vs.* delay (EDP) of 1ML-$HfS_2$ high-performance inverter cells, at various $V_{DD}$ (0.4V to 0.7V), made of $L$ = 5 nm and $L$ = 3 nm stacked DG MOSFETs (5 ribbons/device) and $L$ = 0 nm and $L$ = 3 nm stacked SG $D_2$-FETs (9 ribbons/device). The EDP performance of Si HP inverter cells made of $L$ = 12 nm stacked Si GAA MOSFETs ($t_S$ = 5 nm, 8 wires/device) and $L$ = 5 nm stacked Si DG $D_2$-FETs ($t_S$ = 4 nm, 4 ribbons/device) are also shown for comparison. The inverters are loaded with a 50 contacted-gate-pitch-long metal line.[12] The extrinsic capacitances of the cell layout are also included in the load capacitance. $I_{OFF}$ = 10 nA/µm. $\Delta L$ = 4 nm for the $D_2$-FETs.

**Outlook:**

In summary, we have presented an in-depth study of the essential physics and performance potential of several 2D materials towards sub-10 nm gate length high-performance CMOS. We have argued that using very-advanced atomistic-simulation techniques including electron-phonon scattering, such as the DFT-NEGF technique we have used here, is required to achieve reliable and accurate results for 2D-materials. We have extracted from our simulations relevant 2D-material parameters such as $m_{\mathrm{DoS}}$ and mobilities for 5 TMD n- and p-type materials.

We have benchmarked n- and p-type MOSFETs made of 6 different 2D materials, including the most used Mo and W-based TMDs, $P_4$, as well as, the less explored $HfS_2$ and $ZrS_2$ materials, against Si GAA. Our results demonstrate the interest and better scaling potentials of $HfS_2$, $ZrS_2$ and $WS_2$ for sub-10 nm HP CMOS providing that a high-doping concentration could be used in the source and drain extensions to mitigate source starvation effects. We have further shown that a high-drive current, meeting the stringent IRDS-2031 target down to about $L$ = 6 nm, would be achievable using $HfS_2$. We finally proposed a novel transistor concept, the Dynamically-Doped Field-Effect Transistor, that scales better than its MOSFET counterpart and seems very attractive for the sub-10nm gate-length regime. Used in combination with a high-mobility material such as $HfS_2$, it allows for keeping the stringent ITRS HP on current when scaling down to 1 nm gate length. It further shows very attractive power-delay performance and its EDP performance keeps increasing when ultimately scaling down $L$ to 0 nm using a simpler SG architecture with an ultra-compact design (CGP = 22 nm). The Dynamically-Doped Field-Effect Transistor further addresses the grand-challenge of doping in ultra-scaled devices and 2D material in particular.

## METHODS

### QUANTUM TRANSPORT SOLVER

Our quantum-transport solver, ATOMOS,[19] was specifically developed for high-performance computing and the use of computationally-heavy DFT Hamiltonians. It is written in C++ and uses multi-threaded MPI with various levels of parallelism. Ultimately, any heavy vector-matrix or matrix-matrix operations are performed using BLAS and LAPACK.

ATOMOS core transport solver is a Real-Space NEGF solver based on the recursive-Green's function (RGF) algorithm.[52] For completeness, the equations for retarded ($G^R$), lesser ($G^<$) and greater ($G^>$) Green's functions read:[35]

$$G^R = \left(EI_N - H - \Sigma^R\right)^{-1}, \tag{1}$$

$$G^< = G^R \Sigma^< G^{R\dagger}, \tag{2}$$

$$G^> = G^R - G^{R\dagger} + G^<. \tag{3}$$

$E$ is the scalar energy. $I_N$ the identity matrix, $H$ the device Hamiltonian, and $\Sigma^{R,<}$ the retarded, lesser self-energies that include the interaction terms (e.g. with the semi-infinite leads $\Sigma_C^{R,<}$ and the electron – phonon scattering terms $\Sigma_S^{R,<}$) are matrices of rank $N$, the total number of atoms in the device × the number of orbitals/atoms. We efficiently store $H$ and other $G$ matrices using our dedicated sparse block-matrix class, that we specifically customized for the RGF method.

The contact self energies are computed with the Sancho-Rubio method.[53] Electron-phonon scattering is considered using the self-consistent Born approximation.[26] Assuming the phonons stay in equilibrium, the scattering self-energy may be written as:[35]

$$\Sigma_S^<(r_i, r_j, E) = \int \frac{dq}{(2\pi)^3} e^{iq.(r_i - r_j)} \left|M_q\right|^2 \times (N_q + \tfrac{1}{2} \pm \tfrac{1}{2}) G^<(r_i, r_j, E \pm \hbar\omega_q) \quad (15)$$

where $q$ and $\omega_q$ are the phonon wave vector and the corresponding angular frequency, $\hbar$ is the reduced Plank's constant, $N_q$ is the phonon-occupation number. $M_q$ is the electron-phonon coupling matrix, which depends on the exact scattering mechanism. For TMDs we used the DFT-computed e-ph parameters from ref.[16], for $P_4$ those from ref.[45]. Additional details about the e-ph scattering implementation can also be found in SI section 3. To ensure efficient load-balancing, a master-slave-approach-based dynamic scheduler is used to distribute the various energy-momentum (*e*-*k*) points between the different parallel ranks. For optimally generating the energy points, we rely on a recursive adaptive-grid algorithm,[54] using a trapezoidal integration rule and a global-error estimator.

Similarly, a parallel Poisson solver with its own sparse class is used. To expedite the self-consistent Poisson-NEGF convergence, we employ a predictor-corrector method using the Newton scheme.[55] To predict the carrier changes with respect to potential variations, various semi-classical predictor functions have been implemented. For 2D materials, we can well fit the NEGF data using a 2D-DoS model, i.e., using a Fermi-Dirac integral of order 0 (See Charge and DoS Fitting section below and SI Fig. 4). Additional adaptive-damping methods can be used, if the current and charge convergence criteria are not met within a given number of iterations. The anisotropic dielectric permittivity's are taken from ref.[56]

In this work, we have used the Wannierization technique[57] to express the DFT Hamiltonian in a localized-orbital basis that is compatible with the RGF algorithm. The use of advanced and well-optimized algorithms together with high-performance parallel computing allow for scalable and fast calculations. For the 1ML 2D devices simulated here, using a Real-Space DFT-Hamiltonian with longer-range interactions and dissipative transport, a full $I_D(V_G)$ curve is typically achieved within about an hour on one to a few 100 cores, using the latest generation Intel Xeon CPU.

**DFT-Hamiltonian computation**

The electronic states in the various TMD and BP monolayers are modeled using the density-functional theory (DFT)-based ab-initio tool QUANTUM ESSPRESSO.[21] Both the geometry relaxation and the computation of the electronic structure are performed using the generalized gradient approximation and the optB86b exchange-correlation functional.[22] Spin-orbit coupling was not included. The plane-wave cut-off energy and the Monkhorst-Pack *k*-point grid for the Brillouin-zone integration, that we used for the relaxation and band structure calculations, were selected so that the total energy was well converged. The convergence criteria are set to less than $10^{-3}$ eV/Å for the forces acting on each ion, and a difference smaller than $10^{-3}$ eV for the total-energy variation between two subsequent iterations. A vacuum layer of 25 Å was employed in the DFT simulations to cut off the spurious interactions of the periodic images along the out-of-plane (z-direction, see inset of Fig. 2).

We then transformed the Bloch wavefunctions into maximally-localized Wannier functions, typically centered on the ions, with the wannier90 package.[23] Fig. 2 and SI Fig. 9a demonstrate the validity of our MLWF representation for the case of $WS_2$, $ZrS_2$, $HfS_2$, and $P_4$. ATOMOS uses the resulting supercell information, i.e., MLWF and atom positions, lattice vectors, and the localized Hamiltonian matrix elements, as building blocks to create the full-device atomic structure and Hamiltonian matrix. We kept in the device Hamiltonian, the required Wannier-Hamiltonian longer-range interactions (typically 12 to 15 Å). ATOMOS can further rotate the device geometry to a preferential channel orientation within the 2D layer. We assumed periodic boundary conditions in the width (y-axis) direction. They were modeled with 24 $k_Y$ points.

## Charge and DoS Fitting

From our NEGF simulations, we can extract the electron concentration vs. $E_F - E_C$. As can be seen in SI Fig. 4, this can be well fitted by a 2D-DoS model (using a Fermi-Dirac integral of order 0), from which the conduction band DoS, $N_{2D}$, can be extracted by:

$$n = N_{2D}.\ln\left(1 + e^{\frac{E_F - E_C}{k_B T}}\right) \quad (1)$$

In (1), $E_F$ is the Fermi-level, $E_C$ is the conduction band edge, $k_B$ is the Boltzmann constant and $T$ is the absolute temperature. The NEGF-simulated carrier concentrations *vs.* $E_F - E_C$ curves are extracted from a $L$ = 14 nm device using the averaged values of a cross-section in the middle of the channel under low $V_D$ (typically 1mV) bias condition, while varying $V_G$. Due to non-equilibrium transport, the NEGF Fermi level is only known and well defined at the source and drain contacts. At very low $V_D$, we can, however, safely assume a quasi-linear and close to equilibrium regime and that the Fermi-level value in the middle of the channel is halfway between that of the source, $E_{FS}$, and drain, $E_{FD}$. Assuming an equivalent single-band parabolic effective-mass model, we can further define an equivalent DoS mass, $m_{DoS}$, by:

$$N_{2D} = \frac{m_{DoS}.k_B T}{\pi.\hbar^2} \times \frac{1}{t_S} \qquad (2)$$

Note that this DoS mass capture the DFT computed non-parabolicity of the occupied bands close to the conduction band edge. In (2), $\hbar$ is the reduced Plank constant and $t_S$ is the 2D-film thickness, about 0.6 nm for a monolayer TMD (the exact value we used for each studied monolayer can be found in SI Table 1). $t_S$ is used to convert the 2D density from per-area to per-volume unit. Similarly, for a p-type device, we can extract the valence-band DoS, $N_{2D}$ from the DFT-NEGF-simulated hole concentration vs. $E_V - E_F$ using:

$$p = N_{2D}.\ln\left(1 + e^{\frac{E_V - E_F}{k_B T}}\right) \qquad (3)$$

where $E_V$ is the valence-band edge. Using (2), it is then possible to extract the equivalent hole DoS mass.

ASSOCIATED CONTENT

**Supporting Information** accompanies this paper.

AUTHOR INFORMATION

**Corresponding Author**

Correspondence should be addressed to Aryan Afzalian, imec, Kapeldreef 75, 3001 Leuven, Belgium. Email: Aryan.Afzalian@imec.be

**Author Contributions**

A.A. developed the theory and code, performed the simulations and analysis and wrote the manuscript.

**Data availability**

The data that support the findings of this study are available from the corresponding author upon reasonable request.

**Code availability**

Access to the code that is used in this study is restricted by imec legal policy.

## Competing interests

The author declares no competing interests.

ACKNOWLEDGMENT

Part of the computing resources and services used in this work were provided by the VSC (Flemish Supercomputer Center), funded by the Research Foundation - Flanders (FWO) and the Flemish Government.